\documentclass[a4paper,12pt]{article}
\usepackage{amsmath,amsfonts}
\usepackage{graphicx}
\usepackage{psfrag}

\newcommand{\beq}{\begin{equation}}
\newcommand{\eeq}{\end{equation}}
\newcommand{\matrice}{\begin{pmatrix}}
\newcommand{\ematrice}{\end{pmatrix}}
\newcommand{\bea}{\begin{eqnarray}}
\newcommand{\eea}{\end{eqnarray}}

\begin{document}

{\normalsize \hfill SPIN-07/44}\\
\vspace{-1.5cm}
{\normalsize \hfill ITP-UU-07/58}\\
${}$\\

\begin{center}
\vspace{60pt}
{ \Large \bf The Emergence of Spacetime \\
${}$\\
{\large\bf or}\\
${}$\\
Quantum Gravity on Your Desktop}

\vspace{50pt}

{\sl R. Loll}\footnote{email: r.loll@phys.uu.nl}

\vspace{24pt}

Institute for Theoretical Physics,\\
Utrecht University, \\
Leuvenlaan 4, NL-3584 CE Utrecht, The Netherlands.\\


\vspace{48pt}

\end{center}

\begin{center}
{\bf Abstract}
\end{center}
\noindent

Is there an approach to quantum gravity which is conceptually simple, relies on very few fundamental physical principles and ingredients, emphasizes geometric (as opposed to algebraic) properties, 
comes with a definite numerical approximation scheme, and produces robust results, which go beyond showing mere internal consistency of the formalism?
The answer is a resounding {\it yes}: it is the attempt to construct a nonperturbative theory of quantum gravity, valid on all scales, with the technique of so-called {\it Causal Dynamical Triangulations}. 
Despite its conceptual simplicity, the results obtained up to now are far from trivial.
Most remarkable at this stage is perhaps the fully dynamical {\it emergence} of a classical background (and solution to the Einstein equations) from a nonperturbative sum over geometries, without putting in any preferred geometric background at the outset. In addition, there is concrete evidence for the presence of a fractal spacetime foam on Planckian distance scales. The availability of a computational framework provides built-in reality checks of the approach, whose importance can hardly be overestimated.

\vspace{12pt}

\noindent


\newpage

\section{Quantum gravity: aims and ambitions}

A central aim of any theory attempting to address the problem of quantum gravity is to {\it derive} spacetime as is from an underlying, dynamical quantum principle. By ``spacetime" we mean spacetime together with its geometric properties on {\it all} scales, from the largest, cosmological scales to the smallest scale usually considered by physicists, that of the Planck length $\ell_{\rm Pl}=\sqrt{\hbar G_N/c^3}$. 
``Geometric" here is to be understood in a suitably generalized sense at very short distances, where we expect the classical description of spacetime in terms of a differentiable manifold with a smooth Lorentzian metric $g_{\mu\nu}(x)$ to be no longer adequate. The short-distance structure of a particular model of quantum gravity we will be discussing in what follows is incompatible with such a smooth assignment, but nevertheless possesses more `primitive' metric properties, allowing us to measure - in the sense of quantum theory - certain lengths and volumes. The reason for the expected breakdown of classicality near the Planck scale is the dominance of large quantum fluctuations, as expressed in the non-renormalizability of the perturbative quantization of gravity.   

\begin{figure}[ht]
\centering
\vspace*{13pt}
\includegraphics[bb= 280 320 320 510, scale= .8]{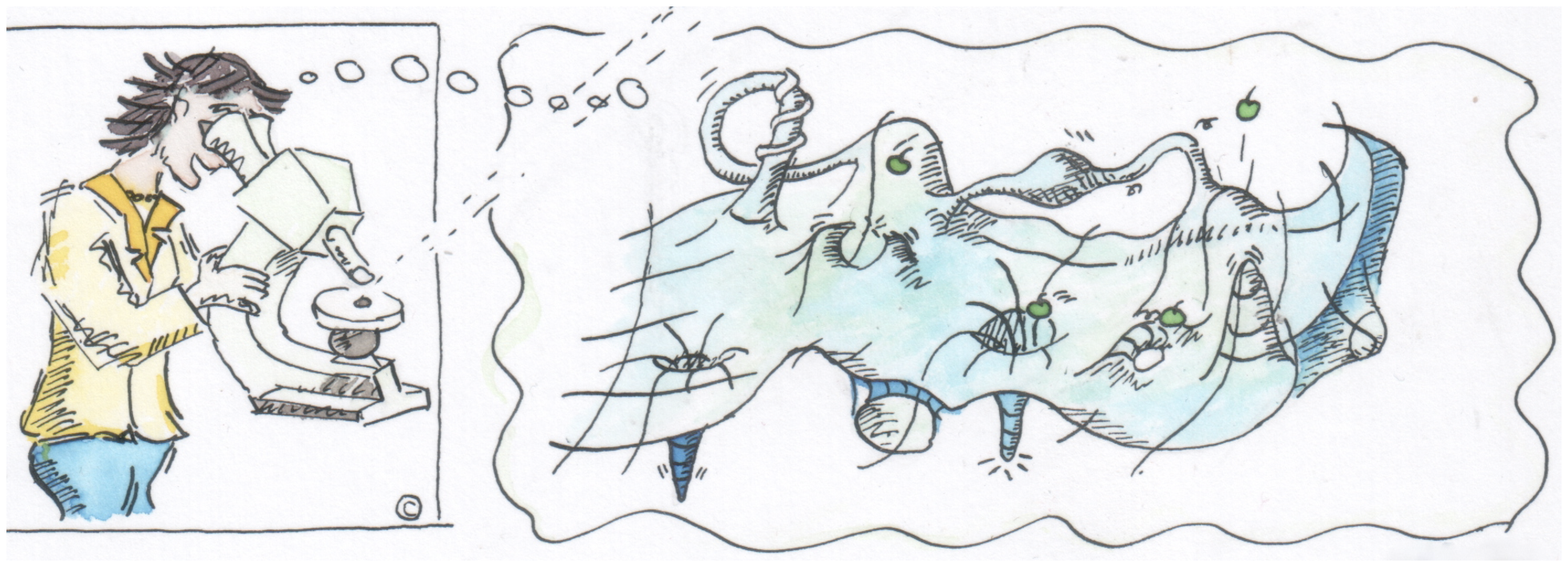}
\vspace*{13pt}
\caption{Practitioner of quantum gravity, studying the Planck-scale microstructure of spacetime
through her ultra-powerful microscope. (Courtesy of E. Rijke.)}
\label{micro2}
\end{figure}

Although qualitative pictures like Wheeler's ``quantum foam" have been around for more than 50 years (c.f. Fig.\ \ref{micro2}), concrete and quantitative models of spacetime at the Planck scale are still lacking.   
One reason why it is so difficult to get a handle on the quantum dynamics of the fluctuations is that they affect what in quantum field theories of the non-gravitational interactions is just part of the fixed background structure, namely, spacetime itself. The failure of the perturbative approach to quantum gravity in terms of linear fluctuations around a fixed background metric implies that the fundamental dynamical degrees of freedom of quantum gravity at the Planck scale are definitely not gravitons. At this stage, we do not yet know what they are. Neither do we have the luxury of hints from experiment or observation of what they might be, although one would certainly hope that given candidate theories of quantum gravity will lead to definite predictions of observable consequences of a nontrivial microstructure of spacetime. Potential examples of this include cumulative effects imprinted on highly energetic photons or neutrinos reaching us from the edge of the universe \cite{light}, and a quantum-gravitational origin of the vacuum energy of empty space, the so-called dark energy.  

Before proceeding, let us recall just how small the Planck length is. With $\ell_{\rm Pl}=10^{-35}m$ it sits at one end of a logarithmic scale reaching all the way to about $10^{24}m$, the extension of the observable universe. Present-day and upcoming high-energy accelerators allow us to probe short distances down to about $10^{-19}m$, which is still a whopping 16 orders of magnitude away from the Planck scale --  far out of reach of any direct observation and presenting a severe handicap to theory builders. With this warning in mind, let us set out our mission statement, which is

\vspace{.5cm}
\fbox{
\begin{minipage}{15cm}
To look for a consistent theory of {\it quantum gravity}, which describes the dynamical behaviour of spacetime geometry on all scales and reproduces Einstein's theory of general relativity on large scales. At the same time, it should also predict new observable phenomena.
\end{minipage}}
\vspace{.5cm}

With all due respect to past efforts and achievements, the search for such a theory has yet to succeed, be it in ``pure gravity" or so-called unified approaches (these days usually based on string theory). Despite occasional claims to the contrary, there is at this stage no compelling evidence whether or not all fundamental interactions have to be unified at the Planck scale. At any rate, it is clear we are dealing with a formidable problem, where technical difficulties often appear entangled with conceptual ones. In the face of these long-standing difficulties, physicists have drawn sometimes far-reaching conclusions about what should be done:

\begin{itemize}
\item Does the theory need new ingredients, either in the form of new symmetries, or in terms of new, fundamental objects? A prime example of the former is (the hitherto unobserved) supersymmetry, and examples of the latter include fundamental strings, loops and higher-dimensional membranes.

\item Do we need to question or modify the fundamental principles of quantum theory and/or general relativity? An example of the former is the suggestion -- motivated by considerations originating in quantum gravity -- that there may be a deterministic theory underlying standard quantum theory \cite{deter}.

\item Most radically, do we need to change our notion of what constitutes a physical theory? Have we reached the end of the road in terms of ``reductionist" physics, despite its past successes in high-energy physics? Do we have to take up ``landscaping" or meta-theorizing?
\end{itemize}   

Many theorists feel a growing unease about these choices. Aren't there any loop holes in the arguments that lead us down these highly speculative roads? Given our lack of knowledge of physics at the Planck scale, wouldn't it be wise to use as few conjectural, untested principles and ingredients as possible? Could it be that the essence of the problem of quantum gravity is `simply' that we need to get a computational handle on an infinite-dimensional, strongly interacting system in a region far from its na\"ive perturbative regime (which is of course difficult)? 

My collaborators and I have been pursuing this latter alternative by treating quantum gravity -- rather prosaically -- as a nonperturbative quantum field theory. Its only (but crucial) difference with other, standard quantum field theories is the fact that it does not rely on any a priori fixed background geometry. 
With the example of nonperturbative QCD in mind, one may already anticipate that this is not possible with analytical methods alone, and this is indeed what we have found. The reason why such a theory may exist is related to the potential presence of a ``real" nonperturbative vacuum, distinct from the na\"ive perturbative vacuum given (in the absence of cosmological constant and matter) by flat Minkowski space. This is related to the old idea of the potential existence in gravity of a non-Gaussian fixed point of the renormalization group in a scenario of ``asymptotic safety" \cite{weinberg,MM,Niedermaier}. The key ideas of our approach to constructing a theory of quantum gravity and the underlying technique of {\it Causal Dynamical Triangulations}, as well as the results achieved so far will be sketched in the following sections.

\section{Making sense of the gravitational path integral}

At the heart of the approach lies an explicit realization of the infamous 
``Sum over Histories", also known as the 
{\it gravitational path integral},
\begin{equation}
Z(G_N,\Lambda)=\int\limits_{\rm spacetime\atop geometries\ g\in\cal G}{\cal D}g\ {\rm e}^{iS^{\rm EH}[g]}, 
\label{gravint}
\end{equation}
with the four-dimensional Einstein-Hilbert action
\begin{equation}
S^{\rm EH}=\frac{1}{G_N}\int d^4x \sqrt{\det g} (R-2 \Lambda),
\end{equation}
where $G_N$ denotes the gravitational coupling or Newton's constant and $\Lambda$ the cosmological constant, and the integral is to be taken over all spacetimes, subject to specified boundary conditions.  
A key point about the expression (\ref{gravint}) is its highly formal nature. Before one has not specified 
the integration space, the integration measure and the conditions under which the integration leads to a meaningful (i.e. non-infinite) result, it should be regarded as a statement of intent rather than a well-defined mathematical quantity. 
Our task is therefore to turn (\ref{gravint}) into a well-defined prescription, evaluate it and show that the final result 
makes sense physically and (hopefully) predicts new physical phenomena. 

The method of causal dynamical triangulations (or CDT, see \cite{cdtrev} for reviews) provides us with a nonperturbative handle to define and 
evaluate 
the path integral $Z(G_N,\Lambda)$ in the presence of a positive cosmological constant $\Lambda$. The
integration space $\cal G$ is a space of causal, Lorentzian geometries, obtained from a certain limiting process, 
which will be described below. In terms of background-independent approaches to quantum gravity, and given the 
time scale of progress in the subject, CDT is still very much a new kid on the block, although it takes its queue 
from earlier, related approaches, most importantly, efforts from the early to mid-90's to quantize gravity in terms
of {\it (Euclidean) dynamical triangulations} \cite{DT-book,4drev}.
Since its inception in 1998 \cite{al}, the CDT method has been implemented and
tested thoroughly, starting in two and three spacetime dimensions, where progress can be made using both
analytical and numerical tools\footnote{Recently, analytic methods have been pushed to derive for the first time a Hamiltonian
in 2+1 dimensional quantum gravity for space-like slices with cylinder topology from a full-fledged sum over
spacetime geometries \cite{blz}.}. These are mostly borrowed from statistical mechanics and the theory of 
critical phenomena in order to evaluate a Wick-rotated\footnote{A Wick rotation is necessary at an
intermediate step to discuss the convergence properties or otherwise of the path integral (\ref{gravint}) and to perform numerical simulations. Applying a Wick rotation to a path integral of Lorentzian spacetimes is {\it not} the same as starting
from a path integral of Euclidean spacetimes \cite{ackl,ajl1,3dcdt}.} 
version of the path integral (\ref{gravint}). Here, we will
present only
results from investigating the physical, four-dimensional theory, the first of which became available in 2004. All
of the central results so far have been obtained with the help of Monte Carlo simulations. 

\begin{figure}[ht]
\centering
\vspace*{13pt}
\includegraphics[width=15cm]{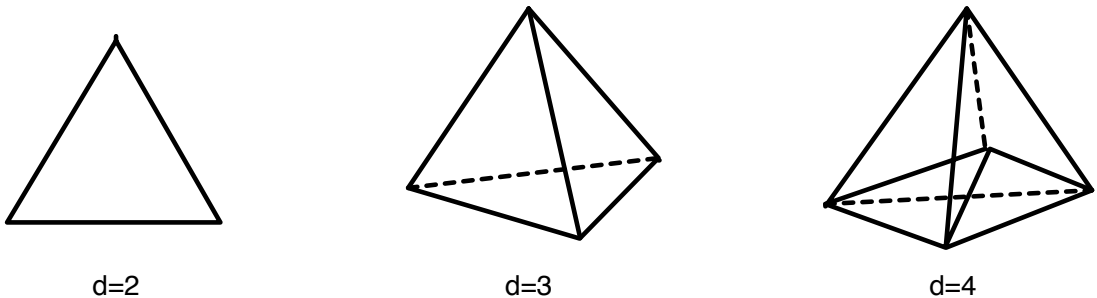}
\vspace*{13pt}
\caption{Triangular building blocks, or $d$-simplices, can be used to construct piecewise flat
$d$-dimensional spaces. From left to right: triangles, tetrahedra and four-simplices.}
\label{buildingblocks}
\end{figure}

How does one go about making sense of the path integral? We use a time-honoured way of
representing the space of all spacetimes, in terms of a set of piecewise flat manifolds in the spirit of Regge's 
``General Relativity without Coordinates" \cite{regge}, which also comes with a ``reggeized" version
$S^{\rm Regge}$ of the Einstein-Hilbert action.
This amounts to constructing regularized versions
${\cal G}_{a,N}$ of the space $\cal G$ of all spacetimes, which are obtained from gluing together $N$
triangular building blocks of typical edge length $a$. They are the four-dimensional analogues of objects which are much easier to visualize, namely,
two-dimensional curved surfaces constructed by gluing together flat triangles (see Fig.\ \ref{buildingblocks}). 
Triangulations of this type are highly versatile and have many applications, for example, in
rendering the shapes of animated figures in Euclidean three-space. However, in our case
there are some important 
differences to keep in mind. The nontrivial curvature properties of geometry we aim to represent in gravitational
applications  -- in keeping with Einstein's theory -- are entirely of an {\it intrinsic} nature and can therefore in principle
be detected by rod-and-clock measurements within the spacetime, without the need to appeal to 
an embedding in a higher-dimensional flat space. Second, classical applications of triangulated 
spaces usually aim to approximate a single, smooth space (which, for example, might be a particular solution to the
classical Einstein equations). By contrast, in quantum-gravitational applications one wants to sample in an
effective manner the space of {\it all} curved spacetimes, which is associated with different requirements,
for example, that of avoiding overcounting. This last point is addressed in CDT by fixing the edge lengths of {\it all}
building blocks (the so-called four-simplices) to the same value, so they are equilateral\footnote{To be precise, in Lorentzian
signature we operate with two types of edges, space- and timelike, associated with two distinct (squared) edge 
lengths. The essential difference with standard triangulations remains in the fact that the edge lengths
cannot assume a continuous set of values.}. After that, only geometrically
distinct gluings of these identical building blocks are considered in the path integral over geometries. Thus, the integration takes place directly on the
space of geometries, without the usual gauge redundancy found in the continuum, where one cannot avoid the
use of coordinates and the accompanying diffeomorphism symmetry.  
 
It would be misleading to say that we wanted to {\it approximate} the space of all spacetimes in this manner. Although it
is clear that we operate at a regularized level, with explicit UV and volume cut-offs $a$ and $N$ present, we have no {\it prior}
knowledge of the abstract space $\cal G$, the relevant integration space for the path integral, or independent ways of deriving it. From comparison with quantum field theory, one would certainly expect it to be highly singular, with the set of smooth, classical configurations corresponding to a set of measure zero. 
The ultimate justification of any particular choice has to come from the concrete results the path integral construction is
able to deliver. 

Finally, we must make an important point regarding the status of this regularized framework. We do {\it not} identify the characteristic edge length $a$ of the simplicial set-up with a minimal, discrete fundamental length scale (equal to the Planck length, say). Rather, we study the
path integral $Z$ in the limit as $a\rightarrow 0$, $N\rightarrow\infty$, which means that individual building blocks are completely
shrunk away. Our concrete realization of the gravitational path integral therefore takes the form
\begin{equation}
\label{discretesum}
Z^{\rm CDT}=\lim_{N\rightarrow\infty\atop a\rightarrow 0} \sum_{{\rm causal,\, triangulated} \atop 
{\rm spacetimes} \, g\in{\cal G}_{a,N}}\frac{1}{C_g}{\rm e}^{iS^{\rm Regge}[g]},
\end{equation}
where $C_g$ denotes the order of the automorphism group of the geometry $g$.\footnote{The measure
used here is the one counting each distinct geometry exactly once, {\it unless} the geometry itself
possesses symmetries (reflected in $C_g>1$), in which case its weight is reduced by a corresponding 
factor.}
In taking this limit, we look for scaling behaviour of physical quantities
indicating the presence of a well-defined continuum limit. By construction, if such a limit exists, the resulting continuum theory 
will {\it not} depend on many of the arbitrarily chosen regularization details, for example, the precise geometry of the
building blocks and the details of the gluing rules. This implies a certain {\it robustness of Planck-scale physics}, as a consequence of the property of {\it universality}, familiar from statistical mechanical systems at a critical point. By contrast, in models for quantum gravity which by postulate or construction are based on some discrete structures at the Planck scale (spatial Wilson loops, causal sets etc.), which are regarded as fundamental, the Planck-scale dynamics will 
generically depend on all the details of how this is done, and therefore be highly non-unique. 
Theoretically speaking such models could still be compatible with the correct classical dynamics on scales larger than the Planck scale (although the proof of this is rather nontrivial, as we will see below), but what we are after in quantum gravity is of course precisely the physics {\it beyond} this 
(semi-)classicality.  
 
\begin{figure}[ht]
\centering
\vspace*{13pt}
\includegraphics[width=12cm]{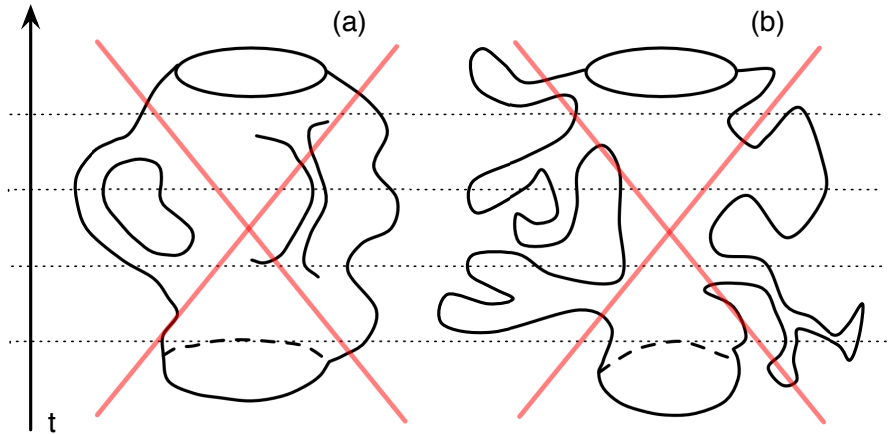}
\vspace*{13pt}
\caption{Forbidden path integral configurations in causal dynamical triangulations. The spacetime topology is fixed, and the sum over histories does not contain a sum over topologies, including, for example, wormhole configurations of the kind depicted in (a). Moreover, the spatial topology is not allowed to change
as a function of time, thus eliminating ``baby universes" and the associated branching and merging points where the light cones and therefore the causal structure are degenerate (b).}
\label{wormsno}
\end{figure}

\section{Universal lessons for nonperturbative quantum gravity}

Having decided that the fundamental building blocks will be equilateral four-simplices, it 
remains to specify the gluing rules according to which they can be put together to
obtain the complete class of piecewise linear spacetimes constituting the 
support of the regularized path integral. A priori, one might worry that there is an infinite number of
valid ways of choosing gluing rules, each one leading to a different theory of quantum gravity,
but this is of course not the case\footnote{Otherwise, we would already be spoiled for choice of
competing quantum gravity theories, whereas in reality, we do not have a single one.}.
It turns out that the combined requirement of having a well-defined path integral 
(after renormalization) and of obtaining a classical limit describing four-dimensional
spacetimes imposes stringent restrictions on the configuration space of the path integral, and thus 
the gluing rules.  
In deriving this result, the availability of numerical techniques for evaluating the sum over 
histories has been absolutely crucial, as will become clear in what follows.

To start with, the gluing rules cannot be too ``liberal", because then the number of allowed
gluings grows too fast as a function of the (discrete) volume $N$ of the triangulations to permit 
us to evaluate the path integral in a well-defined fashion. This happens whenever the number 
of regularized spacetimes $|{\cal G}_{a,N}|$ grows faster than exponentially with $N$, and is
inevitable whenever the ``sum over histories" also subsumes a sum over spacetime
topologies, for example, when wormholes are allowed (Fig.\ \ref{wormsno}(a)). On the other hand, the gluing rules cannot be too restrictive, because we do not
want to kill the local geometrical, curvature degrees of freedom which are the essence of gravity. This leaves us with a ``window of opportunity" where the divergence of the path integral,
$|{\cal G}_{a,N}|\leq {\rm e}^{c N}$ due to the exponential growth of configurations can be renormalized by a suitable choice of a (positive) bare cosmological constant $\lambda_c$, with associated 
Boltzmann weight ${\rm e}^{-\lambda_c N}$ coming from the Euclideanized Einstein-Hilbert action. 

The -- maybe surprising -- conclusion from investigating models inside this window is that they generically are {\it not} associated with a good classical limit in the sense that they do not give rise to
spacetimes which are macroscopically extended and four-dimensional. The point is that in the type of nonperturbative superposition of geometries we are considering, large short-scale fluctuations of the geometry are not suppressed, and will generically lead to a degeneration on larger scales, preventing any {\it emergence} of classical geometry. This finds expression in the fact that the notion of ``dimension", usually regarded as part of the fixed background structure, will in fact become a dynamical quantity in such models, with no guarantee that its value on large scales is equal to 4, although the dimensionality of the building blocks at the cut-off scale $a$ is always 4 by construction! This was first uncovered by simulating models of Euclidean dynamical triangulations, which exhibited the existence of two different phases (depending on the value of the bare Newton constant), one characterized by a Hausdorff dimension $d=2$, and the other by $d=\infty$ \cite{edt1}. The mechanism is very robust, not affected by changing the form of the action or the path integral measure, and thus likely to be present also in other nonperturbative approaches to
quantum gravity.\footnote{A somewhat reminiscent entropic effect (i.e. related to the counting of configurations) is the fact that a generic causal set of discrete volume $N$ has infinite Hausdorff dimension \cite{kr,henson}.}

The only known cure for this undesirable and unphysical behaviour is the one implemented in {\it causal} dynamical triangulations. It consists in restricting the triangulated histories allowed in the sum to those with
a well-behaved causal structure, in the classical sense of forbidding the presence of branching points where the light cone structure is classically singular, as illustrated in Fig.\ \ref{wormsno}(b). This has an immediate physical interpretation: in contrast with the purely Euclidean gravitational path integral advocated by Hawking and others, whose configuration space consists of Riemannian spaces (without any distinction between spatial and time directions and no notion of light cones or causality), causality and the presence of a distinguished future time direction play an important role in constructing the new, causal path integral. Lastly, the imposition of causality constraints by no means entails a generic suppression of large short-scale curvature fluctuations. For example, baby universes which branch off in the spatial directions are not forbidden, and contribute in a crucial way to the nontrivial short-distance structure of the quantum geometry, as we will see later.

The way in which the causality conditions are implemented in the gluing rules for triangular building blocks is by giving the geometries a globally layered structure, labelled by a global, geometrically defined integer-valued proper time $t$.
An added bonus is that this set of Lorentzian piecewise flat geometries comes with a natural ``Wick rotation", i.e. a one-to-one map onto a subset of the space of Euclidean piecewise flat geometries (see \cite{ajl1} for further explanation and  details of implementation). The fact that it is a strict subset leads to a different sum over geometries, which turns out to have a different continuum limit, as already alluded to above. Note that there is  
no way of characterizing this subset intrinsically within the purely Euclidean geometries, because it appeals to causality and the existence of a preferred class of times, both of which are not present in Euclidean gravity.

\section{Quantum gravity on your desktop}

Armed with causal gluing rules for the triangular building blocks, all that remains is to {\it perform} the path sum in
(\ref{discretesum}). 
The beautiful property of the approach at hand is that a well-defined regularization exists, which can be simulated numerically to extract information about the quantum superposition of all spacetimes.\footnote{Contrary to what one finds in perturbative Euclidean path integrals for $d\geq 3$, the problem associated with the conformal factor, which renders the action unbounded below and the path integral ill-defined, is {\it not} present in the nonperturbative CDT path integral \cite{3dcdt,ajl1}, because the divergence is entropically suppressed. In a continuum language, the divergence is cancelled by a nonperturbative Faddeev-Popov determinant,
see \cite{conformal} for a discussion.}

 We employ time-honoured Monte Carlo techniques, adapted to the case of fluctuating lattice geometry, and study the behaviour of the continuum limit via finite-size scaling, extrapolating in a systematic way from the finite, regularized system to an infinite one. 
The only limitation, akin to other quantum field theories studied on the lattice, is computational power, translating into lattice size. This is about as good as it gets in full-fledged, nonperturbative quantum gravity: program your PC or laptop to generate the superposition of geometries, and then perform suitable ``experiments" (that is, measure the expectation values of suitable observables) to determine the geometric properties of the {\it quantum spacetime} obtained in this way. 

As should be clear from the discussion above, in a scheme of this sort one is by no means ensured that what comes out {\it is} the correct theory of quantum gravity. However, the computational handle we have on such models can give us a very good idea of whether we may be on the right track or whether the formulation is fatally flawed. 
This is something that is very difficult to tell if one only has pen and paper at one's disposal to evaluate a sum over such geometries nonperturbatively.

\begin{figure}[ht]
\centering
\vspace*{13pt}
\includegraphics[width=8cm]{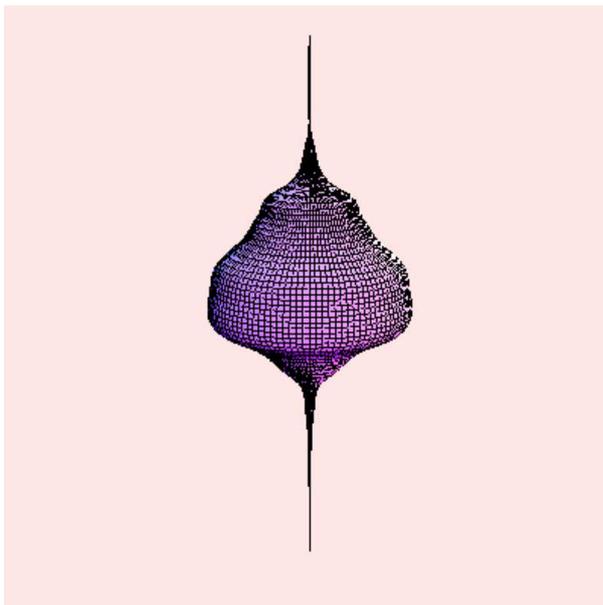}
\vspace*{13pt}
\caption{Emergence of extended geometry: a snapshot from Monte-Carlo simulations of CDT at fixed four-volume $N=91.000$, which shows the spatial three-volume as a function of cosmological proper time (vertical axis).}
\label{unipink}
\end{figure}

The very exciting news is that numerical experiments performed so far on the gravitational path integral defined via CDT have produced concrete evidence of
\begin{itemize}
\item[1.] {\it classical behaviour of the quantum universe on large scales}, as borne out by the existence of a stable, extended quantum geometry which is four-dimensional on large scales, and whose overall shape (Fig.\ \ref{unipink}) is described by a Friedmann cosmology.
\item[2.] Even more intriguingly, we observe strong deviations from classical behaviour on small scales, to the effect that {\it at very short distances, spacetime has a fractal structure and is effectively two-dimensional}.  
\end{itemize}

To give an idea of how such results are obtained, we will look at two distinct ways of measuring the {\it dimensionality} of the quantum spacetime generated by the superposition of geometries. We extract these so-called {\it Hausdorff} and {\it spectral dimensions} from scaling laws which hold in the ensemble. Doing the same on any classical, differentiable manifold, the values of these dimensions will always coincide with the standard (topological) dimension of the manifold. On more general spaces, their values will in general not coincide, nor attain integer values.\footnote{A well-known example is the Sierpinski gasket, a fractal space whose Hausdorff dimension is $D_H\approx 1.585$ and 
spectral dimension is
$D_S\approx 1.365$, independent of scale.}

\subsection{The dimension of quantum spacetime I}

\begin{figure}[ht]
\centering
\vspace*{13pt}
\includegraphics[width=16cm]{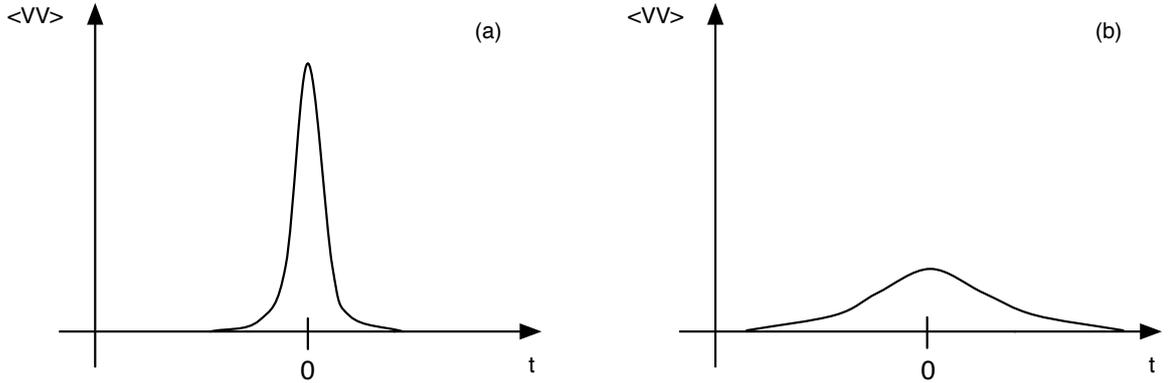}
\vspace*{13pt}
\caption{The qualitative behaviour of the correlator $\langle VV\rangle$ defined in eq.\ (\ref{correl}), for small (a) and large (b) four-volume $V_4$.}
\label{corrsketch}
\end{figure}

One way of extracting an effective dimension of the quantum spacetime is by looking at the behaviour of the correlator
of the three-volume at different discrete times $t$. The three-volume $V_3(t)$ at some (discrete, integer) time $t$ is simply obtained by counting the number of three-simplices or tetrahedra at fixed $t$ (a cut through the four-dimensional triangulation). For simulation-technical reasons we perform our ``experiments" always at fixed four-volume $V_4$\footnote{This implies we are working with the Laplace transform $\tilde Z(G_N,V_4)$ of the original path integral $Z(G_N,\Lambda)$.}, so the relevant correlator is given by
\begin{equation}
\langle VV\rangle (t) :=\langle V_3(0) V_3(t)\rangle_{V_4}=\frac{1}{T} \sum\limits_{s=1}^T 
\langle V_3(s)V_3(s+t )\rangle_{V_4},
\label{correl}
\end{equation} 
where the sum has been taken over all $T$ time steps, and the normalization chosen such that for given volume $N=V_4$
\begin{equation}
\sum\limits_{t=1}^T \langle V_3(0)V_3(t)\rangle_{V_4} =1.
\end{equation}
A qualitative sketch of the behaviour of this correlator as a function of the spacetime volume is given in Fig.\ \ref{corrsketch}. 

\begin{figure}[ht]
\vspace{-3cm}
\centering
\psfrag{tau}{\bf{\large $\tau$}}
\psfrag{VV}{\large\bf $\langle VV \rangle$}
\includegraphics[width=13cm]{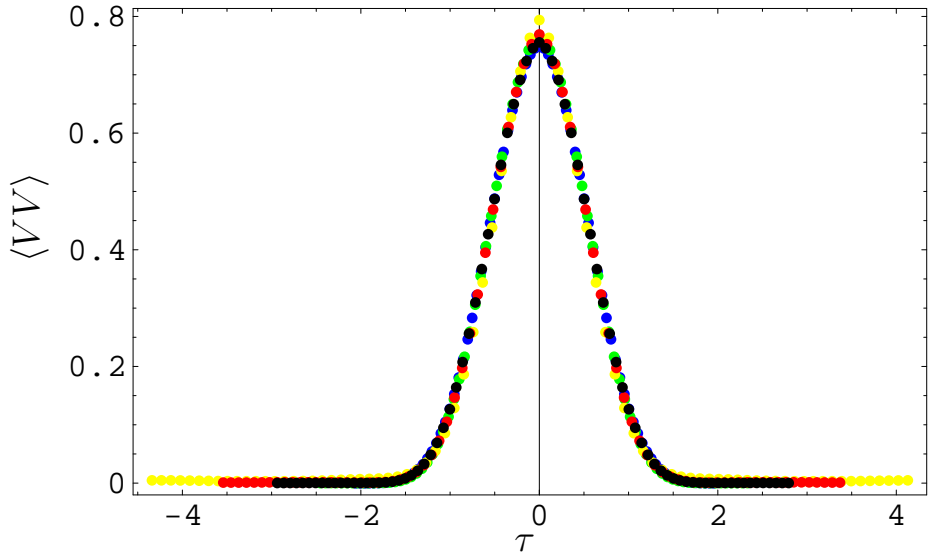}
\vspace{-4.5cm}
\caption{The rescaled correlator $\langle VV\rangle (\tau)$, for spatial volumes
$V_4=$ 22.250, 45.500, 91.000, 181.000 and 362.000, for a universe with $T=80$.}
\label{corrscale}
\end{figure}

In order to understand how $t$ scales as a function of the total volume, we look for a rescaled variable
$\tau=t/V_4^{1/D_H}$ such that the correlators, re-expressed in terms of $\tau$, fall on top of each other for different $V_4$. This is indeed possible, and within measuring accuracy the relevant exponent (the Hausdorff dimension)
has been determined as $D_H =4$ \cite{ajl-prl,ajl-rec}, as one would expect for the scaling of the time function in a four-dimensional space. Fig.\ \ref{corrscale} illustrates the correlators for various volumes after rescaling. The overlap is seen to be excellent, corroborating our assertion of an effective, large-scale dimension of four.

\begin{figure}[ht]
\centering
\vspace*{13pt}
\includegraphics[width=13cm]{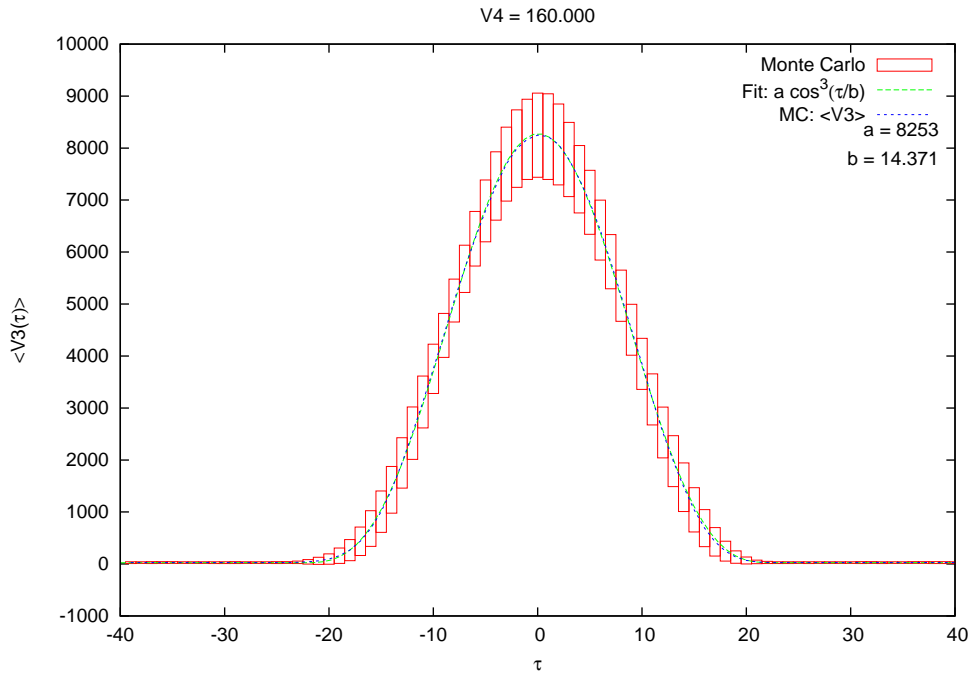}
\vspace*{13pt}
\caption{The shape $\langle V_3(\tau)\rangle$ of the quantum universe, fitted to that of a four-sphere with rescaled proper time, $\langle V_3(\tau)\rangle=a \cos^3(\tau/b)$. Measurements taken for a universe of four-volume $V_4=160.000$ and time extension $T=80$. The fit of the Monte Carlo data to the theoretical curve for the given values of $a$ and $b$ is very good. The vertical boxes quantify the typical fluctuation scale around the expectation value $\langle V_3(\tau)\rangle$.}
\label{spherefit}
\end{figure}

The discussion so far has been about the three-volume of the universe at a given time $t$ or $\tau$. This is of course closely related with the so-called scale factor $a(\tau)$ in cosmology via $a(\tau)\sim \sqrt[3]{V_3(\tau)}$. One may ask two questions: first, what is the ``effective action" for $V_3(\tau)$ obtained from performing the complete path integral over all degrees of freedom other than this global scale? With the
help of the computer simulations, this action has been determined to be a simple minisuperspace action for a homogeneous and isotropic universe \cite{semi}. This is a remarkable result, because it is not obtained from making symmetry assumptions and associated reductions of the configuration space {\it before} the path integral is performed, as is usually the case in discussing quantum cosmologies. On the contrary, here all degrees of freedom have been summed over in the path integral, and their ``collective" effect is seen to lead to just such an action for $V_3(\tau)$ or, equivalently, $a(\tau)$. Second, what are the properties of the particular solution for the quantum universe observed in the simulations? Here one determines its ``typical shape", by which we shall mean the expectation value $\langle V_3(\tau)\rangle$, where the zero of $\tau$ has been defined to coincide with the peak of the curve $V_3(\tau)$. In other words, one averages over configurations of the type illustrated by the snapshot of Fig.\ \ref{unipink}. The result is again remarkable, because this shape can be fitted almost perfectly to that of a metric four-sphere, which is nothing but the solution to the (Euclidean) vacuum Einstein equations in the presence of a positive cosmological constant $\Lambda$. Its line element in proper-time variables can be written as 
\begin{equation}
ds^2=d\tau^2+a^2(\tau) d\Omega^2_{(3)},\;\;\; a(\tau)=\sqrt{\frac{3}{\Lambda}}\cos \sqrt{\frac{\Lambda}{3}}\tau,
\label{lineel}
\end{equation}
where $d\Omega^2_{(3)}$ denotes the line element of the metric on the unit three-sphere. The corresponding fit, Fig.\ \ref{spherefit}, is made after fixing two parameters, namely, the constant proportionality factor between our ``proper time" and the $\tau$ of eq.\ \ref{lineel}, and an overall scale, relating to the total volume of the four-sphere, both of which are expected to be present in our set-up \cite{agjl}. All of this is extremely encouraging, in the sense that a classical $S^4$-geometry -- which moreover is a solution to the classical equations of motion -- is indeed {\it emerging} from a genuinely nonperturbative sum over histories (which treats all geometries ``democratically"), without us having put it in as a background in the first place. While these properties pertain to large-scale, global properties of the quantum universe, the considerations described next also yield information about the short-scale structure of quantum spacetime. 

\subsection{The dimension of quantum spacetime II}

The ``experiment" we are going to set up next can be thought of as a diffusion process in spacetime. We will use the fact that this process is sensitive to the dimension to extract another type of effective dimensionality of the quantum spacetime we have generated. Since the diffusion probes properties of the Laplacian on the space, the most natural interpretation of this so-called spectral dimension is that of the dimension felt by (scalar) matter placed in the spacetime. For reference purposes, recall that the diffusion equation on {\it flat} $d$-dimensional space $M$ is given by
\begin{equation}
\partial_\sigma P(x,\sigma)=\nabla^2 P(x,\sigma),
\label{diffeq}
\end{equation}
where $\sigma$ denotes the (external) diffusion time characterizing the process, and $P(x,\sigma)$ is the probability distribution at a point $x$ after diffusion time $\sigma$. If one assumes that the process is initially peaked in the form of a $\delta$-function at the point $x_0$, the solution to (\ref{diffeq}) is
\begin{equation}
P(x,x_0,\sigma)=\frac{{\rm e}^{-(x-x_0)^2/4\sigma}}{(4\pi\sigma)^{d/2}},
\end{equation}
where attention should be paid to the overall dependence $\sim \sigma^{-d/2}$. This dependence can be extracted in a cleaner way by considering only paths which return back to the initial point $x_0$. To also obtain a diffeomorphism-invariant quantity, we moreover should integrate over all initial points, leading to the average return probability
\begin{equation}
{\cal R}_V(\sigma):=\frac{1}{V(M)} \int_Md^dx\ P(x,x,\sigma)=\frac{1}{(4\pi\sigma)^{d/2}},
\end{equation}
where the last equality again holds for the special case of a flat space $M$ of volume $V$. A similar diffusion process can be set up on an arbitrary curved $d$-dimensional Riemannian space, and leads to the same $d$-dependence for the return probability. The convenient feature for our purposes is the fact that diffusion can be defined on much more general spaces than smooth manifolds, for example, on the triangulated spaces we are using, but also on fractals, say \cite{fractal}. In these cases one is often interested in probing an (a priori unknown) geometry. This means that one studies the behaviour of diffusion, and then extracts a {\it spectral dimension} by looking at the leading behaviour in $\sigma$ of the average return probability. In our case, the diffusion process is defined in terms of a discrete random walker between neighbouring four-simplices, where in each discrete time step there is an equal probability for the walker to hop to one of its five neighbouring four-simplices \cite{spectral}. For the ensemble of CDT geometries, one then determines the spectral
dimension $D_S(\sigma)$ as the logarithmic derivative of the ensemble average
\begin{equation}
D_S(\sigma):=-2\frac{d}{d\log\sigma}\langle{\cal R}(\sigma)\rangle_V, \;\;\;\sigma\leq V^{2/D_S}. 
\end{equation}

\begin{figure}[t]
\psfrag{X}{{$\sigma$}}
\psfrag{Y}{{ $D_S(\sigma)$}}
\centerline{\scalebox{1.2}{\rotatebox{0}{\includegraphics{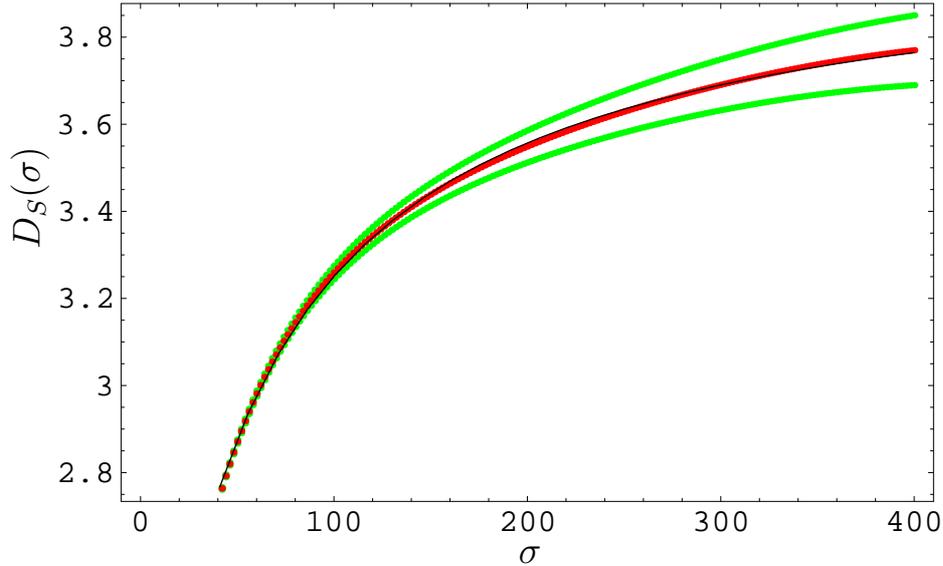}}}}
\caption{The spectral dimension $D_S$ of the quantum universe as
function of the diffusion time $\sigma$, measured at a four-volume $V_4=181.000$.
The central fat curve represents the numerical data, superimposed with a best fit (thin curve -
hardly distinguishable). The two outer
curves represent error bars.}
\label{spectralfinal}
\end{figure}

\begin{figure}[t]
\centerline{\scalebox{1.1}{\rotatebox{0}{\includegraphics{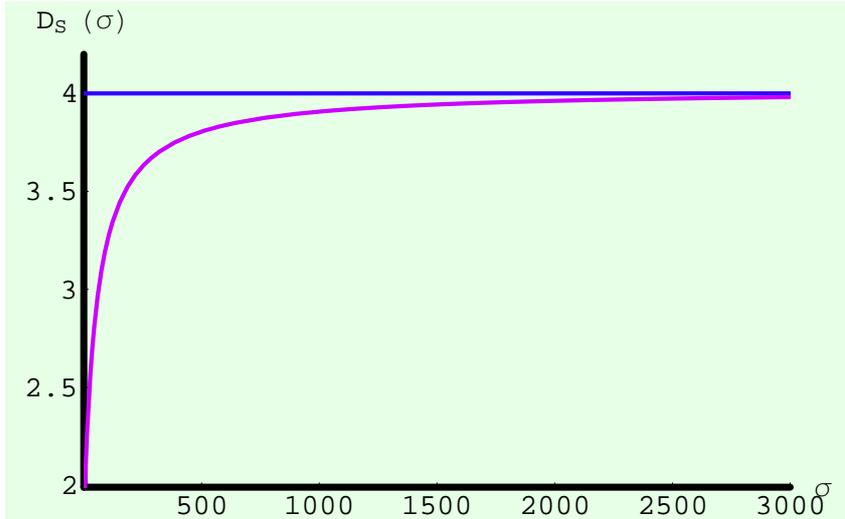}}}}
\caption{The spectral dimension $D_S(\sigma)$ of the CDT-generated quantum universe (lower curve), contrasted
with the corresponding curve for an arbitrary classical spacetime, simply given by the
constant function $D_S(\sigma)=4$.
}
\label{specdimnew}
\end{figure}

As indicated by the formula, the spectral dimension can in general depend on the diffusion time $\sigma$, that is, effectively on the typical length scale $\sim \sqrt{\sigma}$ probed by the diffusion process. The results of
measuring $D_S(\sigma)$ over a range $\sigma\in [40,400]$ are depicted in Fig.\ \ref{spectralfinal}. Rather surprisingly, they indeed display a nontrivial dependence of the spectral dimension on the distance probed!
From a best fit \cite{spectral,ajl-rec}, one extrapolates the asymptotic values
\begin{equation}
D_S(\sigma\rightarrow \infty) = 4.02 \pm 0.1, \;\;\;\;\;\;
D_S(\sigma\rightarrow 0) = 1.82 \pm 0.25.
\end{equation}
We interpret these numbers as reconfirming the four-dimensionality of spacetime on large scales, and 
(tentatively) as a two-dimensionality on short scales. Obviously, the entire distance range where $D_S$ differs significantly from 4 cannot possibly correspond to a classical geometry, and therefore must be an indicator of nontrivial structure at and near the Planck scale. Fig.\ \ref{specdimnew} depicts the extrapolated curve $D_S(\sigma)$, together with the corresponding (straight) curve $D_S(\sigma)=4$ which one would find for a classical manifold. Apart from giving us a quantitative glimpse of the possible structure of Planck-scale ``spacetime foam", this finding is remarkable in that a {\it dynamical dimensional reduction} seems to
be taking place at short distances, which may be associated with an effectively lower-dimensional and therefore more benign behaviour than could have been anticipated. In addition, computer measurements made in the slices of constant time \cite{ajl-rec} indicate the presence of fractal structure on the shortest length scales. This
could be evidence of a new ``paradigm" of geometric structure at and below the Planck scale: there is no fundamental discrete scale corresponding to a Planck scale cut-off, but instead a transition region to a regime where the geometric structure (``geometry" again understood in a generalized sense) becomes fractal and self-similar. One can go to ever smaller, sub-Planckian scales, but nothing interesting happens -- physics continues to look just the same. 

Most intriguingly, there is a completely independent approach to quantum gravity, in terms of a nonperturbative renormalization group analysis, which reaches very similar conclusions, with evidence for both $D_S=2$ and fractality on short scales \cite{erg,MM}. The reasons for this coincidence remain at this stage unclear, although one could of course hope that they provide additional confirmation that we are on the right track in constructing the correct quantum gravity theory.

\section{Understanding gravity plus matter}

Our real, existing universe is not empty, and one way in which one might find evidence for quantum-gravitational effects is through the coupling of gravity to other fields. 
Also, we do not know to what extent non-gravitational degrees of freedom played a role in the ``quantum gravity regime" of the {\it very} early universe. One is therefore interested in studying coupled systems of matter and gravity in various regimes. In the context of the CDT path integral, it is straightforward to
couple matter systems by simply performing a double sum, one over all triangulated spacetimes, and then for each such spacetime over all configurations of the matter system (this could be a discretized scalar or gauge field, just as in regular lattice field theory). Schematically, the combined path integral looks like
\begin{equation}
Z(G_N,\Lambda,\kappa_{\rm Matter})=\int\limits_{\rm spacetime\atop geometries\ g\in\cal G}{\cal D}g
\int\limits_{\rm matter\ configurations\atop \{\phi\}\ on\ g}{\cal D}\phi \
{\rm e}^{iS^{\rm EH+Matter}[g,\phi]}, 
\label{coupleint}
\end{equation}
where $\kappa_{\rm Matter}$ denotes the matter couplings.
These coupled CDT-systems are rather complex, and not even in spacetime dimension two analytic solutions are known.\footnote{See \cite{ising} for recent semi-analytical results for spin systems on 2d CDT lattices.} In addition, they throw up a number of subtleties to do with the scarcity of suitable observables to be measured, the matching with classical solutions, and the role played by the Euclideanization which is necessary to perform the simulations.
\begin{figure}[ht]
\centering
\vspace*{13pt}
\includegraphics[width=8cm]{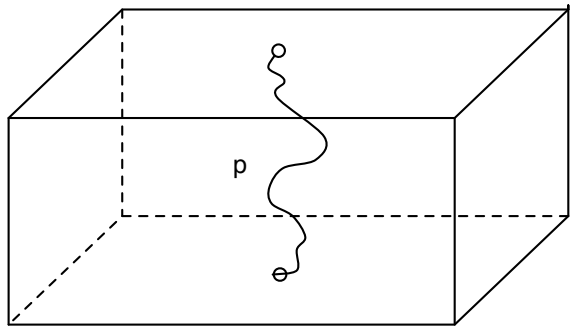}
\vspace*{13pt}
\caption{Worldline $p$ of a single massive particle coupled to pure geometry.}
\label{particlebox}
\end{figure}

The main effort in four dimensions up to now has been on the pure gravity theory, but smaller studies of matter-coupled systems are under way and are essentially of two types. Firstly, simulations of a regularized version of the path integral (\ref{coupleint}) with $\phi$ a scalar field have been performed, with results so far showing little qualitative difference with the pure gravity case. No drastic changes have been observed in the dynamically generated quantum geometry. Secondly, we are studying a situation which may be more closely related to concrete physical observables, namely, a spacetime with one or several heavy pointlike objects, as illustrated in Fig.\ \ref{particlebox}. The Euclidean action corresponding to the situation of one pointlike particle is given by
\begin{equation}
\int {\cal D}g \int {\cal D}p\ {\rm e}^{-S^{EH}(g)-m L(p)},
\label{partact}
\end{equation}  
where $p$ denotes particle paths of length $L(p)$, and $m$ is a mass parameter. In the same way as the nonperturbative pure-gravity path integral gives rise to an emergent geometry which is a four-sphere, one may expect here -- at least for sufficiently large mass $m$ -- to uncover some relation with the Euclidean Schwarzschild-de Sitter solution\footnote{There is an issue of which global boundary conditions are relevant for the computer simulations: 
firstly, note that no choice of a compact time period makes (\ref{ssdes}) into a smooth metric 
everywhere, and secondly, any periodic identification in time with period $\propto M$ -- as usually employed in the
Euclidean pure Schwarzschild case -- 
does not have a good $M\rightarrow 0$ limit.}
\begin{equation}
ds^2=(1-\frac{2 G_N M}{r}-\frac{\Lambda r^2}{3})dt^2+dr^2 \frac{1}{(1-\frac{2 G_N M}{r}-\frac{\Lambda r^2}{3})}+r^2 d\Omega^2_{(2)}.
\label{ssdes}
\end{equation}
In simulations of (\ref{partact}), one does indeed observe a change in spacetime geometry as the mass $m$ is scaled up from zero. Its quantitative aspects and their relation with the geometry (\ref{ssdes}) are currently under investigation. The situation with more than one world line is also of interest, in the first place as a further test of the correct classical limit of the quantum gravity theory defined via CDT, which should reproduce, amongst other properties, the correct classical law for gravitational attraction. In implementing this, one has to deal with the standard difficulty of any nonperturbative approach, namely, the absence of an a priori defined background geometry. In the case at hand, one has to identify suitable, generally covariant observables (and associated computer experiments) which encode the {\it fixed-geometry} set-up of ``a point mass at point $x$ and a point mass at point $y$ attracting each other according to Newton's inverse-square law" in a completely nonperturbative setting. This remains a challenging and nontrivial task in any approach to quantum gravity.

\begin{figure}[ht]
\centering
\vspace*{13pt}
\includegraphics[width=15cm]{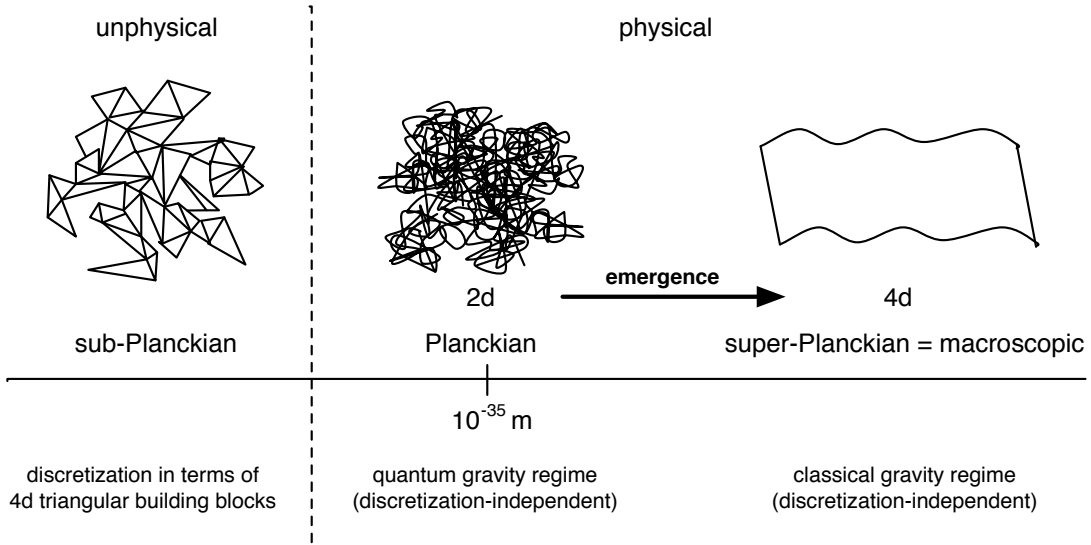}
\vspace*{13pt}
\caption{Causal dynamical triangulations at a glance: from an unphysical, regularized starting point (left of dashed line), we obtain by a nontrivial limiting process the theory which hopefully {\it is} the correct theory of quantum gravity on all scales (right of dashed line), and which in turn consists of a highly non-classical quantum regime around the Planck scale, and an emergent, extended background geometry on large, macroscopic scales.}
\label{emergence}
\end{figure}

\section{Conclusion}

Fig.\ \ref{emergence} summarizes the situation described by causal dynamical triangulations. Starting point (left of the dashed line) is the regularized form of the sum over geometries in terms of causal triangulations, which in itself is unphysical. By taking the continuum limit of this formulation (achieved by fine-tuning the bare cosmological constant to its critical value \cite{ajl-rec}), one arrives at a continuum theory of quantum gravity. In practice, this is the regime of the computer simulations starting at a typical scale $\ell\gg a$, where $a$ is the length scale of individual triangular building blocks (which in an analytic treatment would be sent to zero). This physical continuum theory possesses two different regimes: a quantum regime at the Planck scale, providing a concrete realization of ``spacetime foam", with fractal properties and a spectral dimension $D_S=2$, which then with increasing length scale smoothly and rather rapidly goes over to an effectively classical regime. 

In a beautiful illustration of the phenomenon of emergence, this classical regime turns out to be an extended four-dimensional geometry, which on large scales and in the absence of matter is that of de Sitter space. By universality, the emergent quantum spacetime and the underlying physical theory obtained in this manner do not depend on details of the discrete, sub-Planckian formulation (type of building blocks, action, measure, ...). Their geometric and physical properties are at this stage only partially understood and the subject of ongoing investigations. On the one hand, one wants to verify further details of the classical limit of CDT's quantum geometry, and understand whether they agree with the behaviour expected from general relativity, with and without matter. On the other hand, one's primary interest is obviously in uncovering the detailed structure of the new {\it quantum} regime,
which so far has not been described by other candidate theories of quantum gravity, and to relate it to observable consequences on macroscopic scales.

Key in these considerations is always the identification of suitable observables which are both adapted to the background-free formulation and
can be measured with sufficient accuracy on the (after all relatively small) lattices under consideration, with a typical size of several hundred thousand building blocks. Returning to a point made in the introduction, the crucial tool which has enabled us to make any progress at all in understanding the properties and validity of the proposed quantum gravity theory is the applicability of  
nonperturbative lattice methods. Adapted to the case of variable lattice geometry, and combined with elements of the theory of critical systems, they have given us access to the elusive strongly coupled regime of quantum gravity, and led to unprecedented and concrete results. These findings are very encouraging, but much remains to be explored - stay tuned for more!

\vspace{.7cm}

\noindent {\bf Acknowledgements.} The author thanks J. Ambj\o rn and J. Jurkiewicz for the fruitful collaboration which led to many of the results described in this article.
She also acknowledges support through the European Network on Random Geometry
ENRAGE, contract MRTN-CT-2004-005616, as well as by the 
Netherlands Organisation for Scientific Research (NWO) under their VICI program.

\bibliographystyle{unsrt}

\end{document}